# Convolution Neural Network Model Framework to Predict Microscale Drag Force for Turbulent Flow in Porous Media

Vishal Srikanth[1] and Andrey V. Kuznetsov[1]†

[1]Department of Mechanical and Aerospace Engineering, North Carolina State University, Raleigh, NC 27695, USA

Convolution Neural Networks (CNN) are well-suited to model the nonlinear relationship between the microscale geometry of porous media and the corresponding flow distribution, thereby accurately and efficiently coupling the flow behavior at the micro- and macro- scale levels. In this paper, we have identified the challenges involved in implementing CNNs for macroscale model closure in the turbulent flow regime, particularly in the prediction of the drag force components arising from the microscale level. We report that significant error is incurred in the crucial data preparation step when the Reynolds averaged pressure and velocity distributions are interpolated from unstructured stretched grids used for Large Eddy Simulation (LES) to the structured uniform grids used by the CNN model. We show that the range of the microscale velocity values is 10 times larger than the range of the pressure values. This invalidates the use of the mean squared error loss function to train the CNN model for multivariate prediction. We have developed a CNN model framework that addresses these challenges by proposing a conservative interpolation method and a normalized mean squared error loss function. We simulated a model dataset to train the CNN for turbulent flow prediction in periodic porous media composed of cylindrical solid obstacles with square cross-section by varying the porosity in the range 0.3 to 0.88. We demonstrate that the resulting CNN model predicts the pressure and viscous drag forces with less than 10% mean absolute error when compared to LES while offering a speedup of $O(10^6)$.

## 1. Introduction

Turbulent flow is encountered in porous media in numerous critical applications: ranging from the design of efficient heat exchangers (Jiang *et al.* 2001), coatings to reduce aerodynamic drag and noise (Gómez-De-Segura and García-Mayoral 2019; Zhou *et al.* 2018), optimization of wind turbine performance (Zamani *et al.* 2021), to the prediction of the spread of forest fires (Mell *et al.* 2009). The turbulent flow distribution at the microscale level (pore scale) is inhomogeneous and exhibits a nonlinear relationship with the solid obstacle geometry and the flow Reynolds number (Huang *et al.* 2022; Srikanth *et al.* 2021b). Consequently, the prediction of the macroscale properties of turbulent flow in porous media, such as the drag force and the surface averaged Nusselt number, becomes challenging. This challenge is compounded by the fact that there are infinitely many possible solid obstacle geometries that form the porous matrix. Therefore, it is not practical to rely on empirical data to model the flow behavior. The lack of efficient and robust closure models for macroscale porous media flow inhibits the discovery of optimized porous geometries and introduces substantial error during prediction in the aforementioned applications. The goal of the present study is to develop an efficient data driven model framework to learn the relationship between the geometry and the flow distribution at the microscale level, and use it to predict macroscale turbulent flow properties.

In the turbulent flow regime, the drag force is one of the dominant components of the transport of momentum in porous media (Jin and Kuznetsov 2017). Several empirical models of the drag force were developed based on the concept of permeability – a solely geometric parameter that describes the pressure drop across the porous medium. However, in turbulent flow, nonlinearities arise in the dependence of the drag on the Reynolds number such that the Darcy-Dupuit-Forchheimer model becomes invalid. Taylor series expansion of the Forchheimer coefficient with respect to the flow velocity is suggested to model the nonlinearity of the drag with the introduction of additional model constants (Rao *et al.* 2020). However, the application of the Darcy-Dupuit-Forchheimer model in the turbulent flow regime is a curve-fitting procedure that does not include the turbulent flow physics that causes the drag force. Additionally, the permeability and Forchheimer coefficients are unique to the solid obstacle geometry and empirical relations of these model constants are available for certain commonly occurring solid obstacle geometries like circular cylinders or packed bed of spheres. To the best of the authors' knowledge, a universal relation for these macroscale model constants encompassing a wide range of solid obstacle geometries does not exist.

† Email address for correspondence: avkuznet@ncsu.edu

Neural networks are well-suited to combine the flow behaviors observed for different solid obstacle geometries into a single, unified model without sacrificing accuracy and efficiency. In fluid dynamics, neural networks have been used for reduced-order modeling, flow analysis, and the discovery of governing equations (Duraisamy *et al.* 2019). Previous work on the use of neural networks to model flow in porous media has focused on the prediction of laminar flows. There are two types of neural network models in the literature that are developed for porous media flows: (1) models that predict the permeability of the porous medium at the macroscale level, and (2) models that predict the flow distribution at the microscale level.

El-Tabach *et al.* (2014) developed an artificial neural network model for the pressure drop across a metallic porous medium as a function of the inlet pressure, mass flow rate, and temperature without considering the possibility of varying the solid obstacle geometry. To consider the variation of solid obstacle geometry, morphological descriptors like the pore size distribution, porosity, and number of pores are often provided as inputs to the neural network model of permeability (Liu *et al.* 2022; Wang *et al.* 2020). However, these descriptors do not account for the shape of the solid obstacle, which is an important consideration in turbulent flow (Chu *et al.* 2018; He *et al.* 2019; Srikanth *et al.* 2021b). Several researchers have used geometric functions such as the phase volume fraction and distance functions as input to the neural network so that arbitrary solid obstacle shapes can be considered (Graczyk and Matyka 2020; Ramos *et al.* 2023; Wu *et al.* 2018; Zhang *et al.* 2022). The idea is to represent these functions on a discrete set of points forming a 2D or 3D tensor and provide this as an input to the neural network, which then learns the solution map of the microscale geometry and macroscale permeability. The limitation of this approach is that the geometry and the flow solution are of different scales – microscale and macroscale. While it is theoretically possible to achieve deep learning of the relationship between geometry and flow across different scales, the model training requires a large number of samples. For example, between 10,000 – 100,000 samples were used in Graczyk and Matyka (2020), Ramos *et al.* (2023), Wu *et al.* (2018), and Zhang *et al.* (2022), which was feasible because laminar flow was considered. However, this approach is unsuitable for turbulent flow in porous media since generating a large number of samples is prohibitively computationally expensive.

An alternative approach is to train the neural network model to learn the underlying spatial relationship between the solid obstacle geometry and the flow distribution to accurately predict the flow features that give rise to the drag force. Ideally, the pressure and shear stress distributions on the solid obstacle surface should be modeled by the neural network since the drag force is evaluated on the surface of the solid obstacle. Even though direct prediction of surface stresses was successfully demonstrated for flow around airfoils (Zhu *et al.* 2019), it is challenging to represent the solid obstacle surface topology in porous media since there are numerous solid obstacle shapes to consider. For example, the surface topology for square solid obstacles is different from that of circular solid obstacles due to the presence of sharp vertices in the square geometry that leads to local maxima in the pressure distribution. There is no straightforward approach to convey the presence of features like sharp vertices or curvature to the neural network model.

Therefore, the neural network model must first predict the flow distribution in the entire pore volume and then use the flow distribution to evaluate the drag forces. Such models have been developed for single and multiphase laminar flows in porous media by using autoencoder convolution neural network architectures (Feng and Huang 2020; Marcato *et al.* 2023; Santos *et al.* 2020; Takbiri-Borujeni *et al.* 2020; Wang *et al.*

2021). Feng and Huang (2020) developed a 2D model to predict the laminar two-phase flow interface position and pressure distribution inside a random pore geometry. Santos *et al.* (2020) developed a 3D Convolutional Neural Network (CNN) architecture to predict the single phase laminar flow velocity distribution by providing the geometry of porous rocks as input. Marcato *et al.* (2023) applied a similar CNN architecture to predict the local concentration field for solute transport in porous media. Both Marcato *et al.* (2023) and Santos *et al.* (2020) suggest the inclusion of 4 geometrical input features to train the model more effectively for random porous geometries: Euclidean distance function, Cartesian position, tortuosity, and pore thickness. Since the autoencoder CNN model learns the direct relationship between the microscale geometry and the corresponding flow distribution, fewer samples are required to train the model when compared to CNN models that directly predict permeability. The number of samples used to train the autoencoder CNN models in Feng and Huang (2020), Marcato *et al.* (2023), Santos *et al.* (2020), Takbiri-Borujeni *et al.* (2020), and Wang *et al.* (2021) is between 200-1,000, which is at least 1 order of magnitude less than the number of samples used for the CNN models that directly predict the permeability.

While these neural network models have demonstrated success in predicting laminar flow velocity distributions, turbulent flow prediction presents some challenges. Turbulent flow is inherently unsteady and it requires Reynolds averaging so that the statistically steady solution can be modeled. Reynolds averaging of the turbulent flow is an acceptable approximation since the drag force can be calculated from the Reynolds averaged flow solution. The temporal variation of the drag force is assumed to be negligible for the following reasoning. The dynamics of the microscale turbulent structures inside the pores has a significant influence on the pressure and shear stress distribution on the solid obstacle surface (Huang *et al.* 2022), which is included in the Reynolds averaged flow distribution. In practical applications, the length and time scales of the microscale turbulent structures will be much smaller than the macroscale turbulent structures that are formed in an equivalent clear fluid domain. This is caused by the pore scale suppression of turbulence in porous media since macroscale turbulent structures do not survive inside porous media at the microscale level for values of porosity less than 0.95 (Rao and Jin 2022). Additionally, the phase difference in the microscale turbulent vortex dynamics behind the individual solid obstacles substantially decreases the amplitude of the temporal fluctuation of drag force after volume-averaging over the representative elementary volume (REV) of the porous medium (Srikanth *et al.* 2021b). Therefore, we assume that the small scale of the temporal variation of the drag force inside porous media will result in a negligible influence on macroscale turbulence transport and consider only the Reynolds averaged flow in our model.

CNN models have been used by other researchers to model classical turbulent wall bounded flows (Bhatnagar *et al.* 2019; Kim and Lee 2020; Zhu *et al.* 2019) with a low magnitude of the averaged error over the entire fluid volume. However, a major challenge in modeling the Reynolds averaged turbulent flow inside porous media is that the flow is characterized by thin boundary layers and high shear stress at the solid obstacle surface when compared to laminar flows. In our previous work (Srikanth *et al.* 2021a), we demonstrated that autoencoder CNN models can predict the Reynolds averaged turbulent flow distributions in porous media at the microscale level with low Mean Squared Error (MSE) over the entire pore volume. However, CNN model error is localized in the shear layers surrounding the solid obstacle surface (Srikanth *et al.* 2021a). This compromises the CNN model's ability to predict the drag force since drag is evaluated at the solid obstacle surface. The drag force is one of the dominant closure terms in the double-averaged momentum equation for flow in porous media. In periodic porous media, pressure and viscous drag forces are the only components of the macroscale momentum equation that arise from the microscale level and balance the macroscale pressure gradient in the porous medium. Modeling the drag force at the macroscale level presents significant difficulty since it is not straightforward to accurately estimate the pressure and shear stresses acting on the solid obstacle surface from the macroscale flow velocity and pressure. Therefore, the objective of the present work is to develop a CNN model framework for macroscale model closure of turbulent flow in periodic porous media. First, we have developed an autoencoder CNN model for multivariate prediction of both pressure and velocity distributions. Next, we have developed a conservative interpolation method to minimize the error in the prediction of the pressure and viscous drag forces. We have described the numerical method used to simulate the turbulent flow dataset and the CNN model architecture in section 2. We have discussed the shortcomings of previously used CNN models in predicting turbulent drag forces, proposed solutions to improve model accuracy, and demonstrated the CNN model capabilities with our current approach in section 3.

## 2. Solution Methodology

The development of the data-driven CNN model of turbulent flow in porous media requires two steps: (1) the gathering of a dataset consisting of samples of the solid obstacle geometry and the corresponding turbulent flow distribution, and (2) the optimization of the neural network parameters to fit the turbulent flow dataset.

### *2.1 Turbulent flow dataset*

In the present work, the goal is to develop a CNN model framework to predict the components of the drag force that act on the solid obstacle surface for turbulent flow in porous media. Our objective is to identify the challenges hindering the accurate prediction of the drag force when using the CNN model architecture and propose solutions to overcome them. Therefore, we choose a model problem by considering the turbulent flow in periodic porous media consisting of an in-line arrangement of cylindrical obstacles of square cross-section (figure 1(a)). We have applied periodic boundary conditions at all of the boundaries of the REV and the no-slip boundary condition at the solid obstacle walls. We sustain flow in the periodic domain by applying a volumetric momentum source term that acts as the applied pressure gradient across the REV (Appendix A). We have non-dimensionalized all of the length scales in the geometry with respect to the hydraulic diameter ($d$) of the solid obstacle. In our previous work (Srikanth *et al.* 2021a), we considered cylindrical obstacles of circular cross-section, which presented a challenge due to the occurrence of a symmetry-breaking phenomenon (Srikanth *et al.* 2021b). The development of a CNN model to predict the numerous unique modes of symmetry-breaking is beyond the scope of this work. To circumvent this challenge, we use cylindrical obstacles of square cross-section in the present work since the sharp vertices of the square solid obstacles prevent flow symmetry breaking.

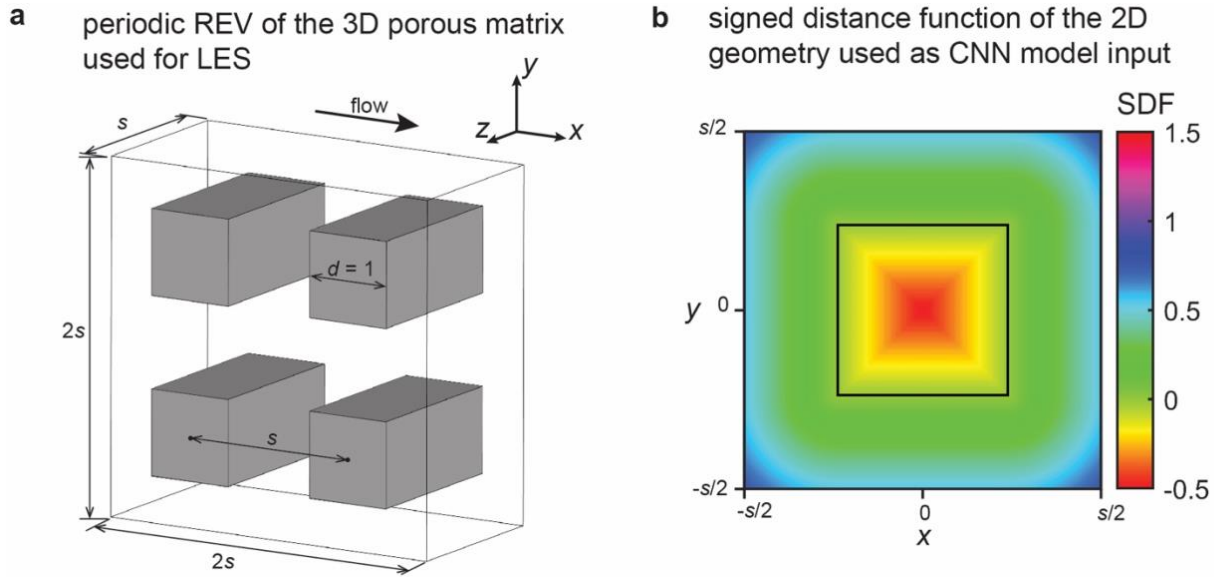

Figure 1: (a) A periodic REV consisting of cylindrical obstacles of square cross-section is used to simulate turbulent flow in porous media with LES turbulence modeling. (b) The microscale geometry of the porous medium is represented in 2D using the signed distance function (SDF) as the input to the CNN model.

We have varied the geometry of the porous medium by changing the porosity (φ) since porosity introduces significant variation in the flow behavior resulting in multiple flow regimes (Srikanth *et al.* 2021b). We will evaluate the success of the CNN model framework based on the ability to predict the different flow behavior experienced in the range of porosities from φ = 0.3 to φ = 0.88.

Data-driven approaches for modeling incompressible Newtonian fluid flow, such as the models discussed in section 1, rely heavily on numerical simulation of the flow dataset. This is because numerical simulation offers superior spatial resolution of the flow distribution at a fraction of the cost of experimental methods. This is especially true for microscale flow in porous media due to the added complexity of inserting probes in the confined pore space or the lack of optical access for particle image velocimetry. In the present work, we numerically simulate the turbulent flow inside the porous medium at the microscale level by using Large Eddy Simulation (LES) with subgrid scale modeling. We have validated the numerical model with experimental data for turbulent flow in an in-line tube bank and performed grid resolution studies to demonstrate that the relevant turbulent scales are resolved in the LES. We have presented the details of the LES model, the validation study, and the grid resolution study in Appendix A.

We have simulated microscale turbulent flow in REV using LES for 61 values of the porosity by varying the distance between the solid obstacles (*s*). The porosity is calculated using equation 2.1, where *CID* is the integer case number ranging from 1 to 61. The Reynolds number of the flow (equation 2.2) is maintained at a constant value of 1000, where $u_m$ is the superficially averaged *x*- velocity and ν is the kinematic viscosity of the fluid.

$$\varphi = 1 - \frac{1}{(1.2+0.03(CID-1))^2} \qquad (2.1)$$

$$Re = \frac{u_m d}{\nu} \qquad (2.2)$$

We have compiled a turbulent flow dataset for this model problem consisting of the microscale distributions for the following flow variables stored at the nodes of the computational grid used for LES: (1) Node position in the Cartesian coordinate system and node connectivity, (2) velocity vector, and (3) pressure.

### *2.2 Neural network model architecture*

The objective of the present work is to develop a neural network approach for closure of macroscale momentum transport in porous media that does not operate like a black-box approach. Rather, the neural network model must learn to predict a physically relevant output from the dataset, such as the turbulent flow velocity, pressure distribution, or momentum sources of pressure and viscous drag. Therefore, we have designed an autoencoder CNN model to take the microscale geometry as the input and predict the corresponding microscale flow distribution as the output. The microscale geometry of the porous medium indicates the locations of the pore space (fluid phase) and the solid obstacles (solid phase). In the present work, the microscale geometry is represented by the signed distance function (SDF), which measures the orthogonal distance between a given point and the nearest solid obstacle surface. The sign of the SDF is positive in the fluid phase and negative in the solid phase. The surface of the solid obstacles are implicitly represented by the isosurface of SDF = 0 (figure 1(b)). The SDF is a more straightforward method to represent the solid obstacle geometry when compared to the volume fraction of the phase even though it requires an additional step in data preparation. This is because the solid obstacle is "immersed" in the computational domain, which results in the possibility of partial solid/fluid volume fraction in some of the grid cells. The presence of grid cells with partial volume fraction introduces the grid dependence of the input, whereas the SDF representation is always grid independent.

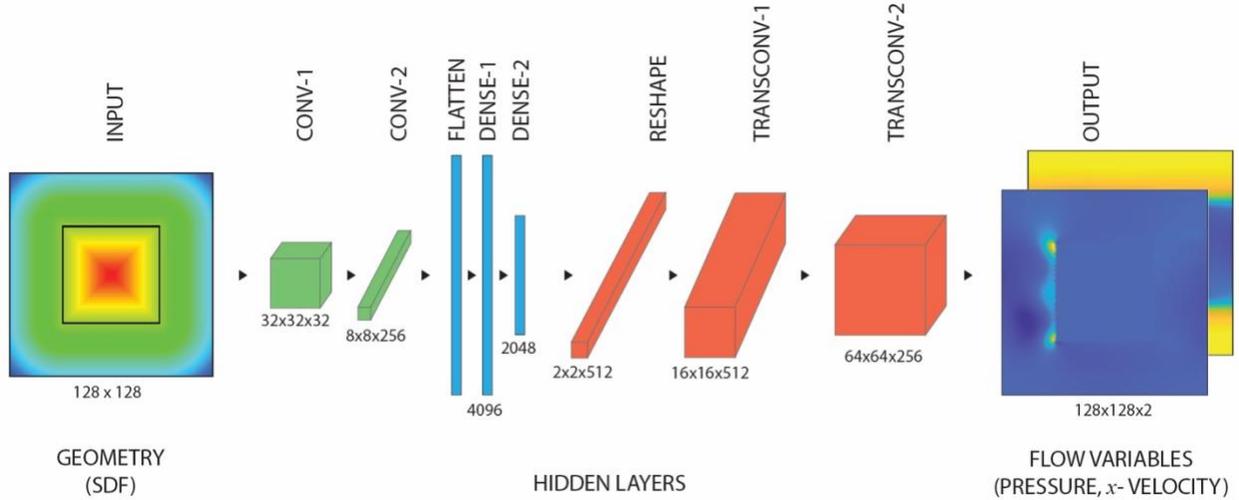

Figure 2: The CNN model maps the geometry of the porous medium to the flow variables through the hidden layers. CONV refers to a 2D convolution layer and TRANSCONV refers to a 2D transpose convolution layer.

The input and output of the CNN model are represented on structured (Cartesian) grids. Although unstructured mesh-based grids will approximate the near-wall pressure and velocity gradient distributions more accurately than Cartesian grids, the mesh will change from case-to-case when the porosity is varied, which is not compatible with the design of the CNN model. Interpolation of the flow variables from the unstructured LES grid to the structured CNN grid presents a challenge, which is discussed in detail in section 3. In the model problem, we have simulated turbulent flow by using a 3D computational domain and then Reynolds averaged the turbulent flow distribution. Since we have assumed that the cross-section of the solid obstacles does not change in the $z$- direction (along the direction of the axis of the cylindrical geometry), the Reynolds averaged turbulent flow statistics are invariant in the $z$- direction. Therefore, it is sufficient to model a two-dimensional Reynolds averaged flow distribution in the $x$- and $y$- directions even though the LES was performed in a three-dimensional computational domain. We have taken the 2D Reynolds averaged flow distribution at the midplane ($z = 0$) of the computational domain to prepare the turbulent flow dataset for the CNN model. We make this approximation to reduce the computational effort while training the CNN model since the focus of the present work is not to consider numerous solid obstacle variations, but to develop the CNN methodology and address implementation challenges in the accurate prediction of the microscale drag force. We note that the CNN model framework can be readily extended to three-dimensional Reynolds averaged flow distributions by including 3D convolutional layers as shown in Santos *et al.* (2020).

The architecture of the autoencoder CNN model used in the present work is as follows (figure 2). The input layer is a tensor of size 128x128. The input layer is followed by 2 convolution layers with 32 (8x8 kernel) and 256 (4x4 kernel) filters, respectively. The convolution layers are followed by 3 dense layers of sizes 4096, 4096, and 2048, respectively. The dense layers are followed by 3 transpose convolution layers with 512 (8x8 kernel), 256 (4x4 kernel), and 32 (2x2 kernel) filters, respectively. Transpose convolution layers are followed by an output layer of shape 128x128x$n_{output}$ to predict $n_{output}$ number of flow variables, each of which is stored on a 128x128 tensor. The hyperparameters: number of convolution/transposed convolution layers and the filter parameters, are selected based on model fit resulting in low magnitude of loss. Hyperparameter optimization is not crucial for the objectives of the present work as long as the error is minimized. The turbulent flow dataset consisting of 61 cases is split into 46 training cases and 15 validation cases. Since there is only a small number of sample cases, the randomization of the cases is performed in batches to avoid the issue of concentration of training cases in a small range of porosity. For example, if the training cases are concentrated in the middle of the porosity range, the model will not perform satisfactorily outside of this training range. Batch randomization is necessary to train neural networks using a small dataset. The activation function used in the present model for the neurons (perceptron) is the ReLu function (Glorot *et al.* 2011). Linear activation is used in the output layer to estimate the numeric values of the flow variables. The mean squared error and normalized mean squared error loss functions (section 3.2) are used to optimize the model weights using the Adam optimization algorithm (Kingma and Ba 2017).

### *2.3 Governing equations of macroscale momentum transport*

The macroscale momentum transport equation (equations 2.3-2.4) for turbulent flow in porous media is derived from the Navier-Stokes equations by applying the double averaging (time and volume) procedure (de Lemos 2012).

$$\rho \left[ \frac{\partial}{\partial t} (\varphi \langle \overline{u_i} \rangle^i) + \frac{\partial}{\partial x_j} (\varphi \langle \overline{u_i u_j} \rangle^i) \right] = -\frac{\partial}{\partial x_i} (\varphi \langle p \rangle^i) + \varphi \frac{\partial}{\partial x_j} \left[ \tau_{ij} - \rho \langle \overline{u'_i u'_j} \rangle^i \right] + \rho \varphi g_i + \overline{R_i} \qquad (2.3)$$

$$\overline{R_i} = \frac{\mu}{\Delta V} \int_{A_{interface}} n_j \, \partial_j \overline{u}_i \, dS - \frac{1}{\Delta V} \int_{A_{interface}} n_i \overline{p} \, dS \qquad (2.4)$$

where $u$ is the microscale flow velocity, $p$ is the microscale pressure, $\tau$ is the macroscale viscous stress tensor, $g$ is the applied pressure gradient, $\rho$ is the fluid density, $\mu$ is the dynamic viscosity of the fluid, $\Delta V$ is the volume of the REV, and $A_{interface}$ is the solid obstacle surface area inside the REV. The angular brackets $\langle \rangle^i$ denote volume averaging, the overbar denotes Reynolds averaging, and the prime denotes the fluctuating component after Reynolds averaging. In the model problem (figure 1), we consider fully developed turbulent flow in porous media consisting of a homogenous periodic arrangement of solid obstacles (by homogeneous, we mean that the REV geometry is copied across the entire porous medium). Therefore, the time and spatial derivatives of the double averaged terms will become equal to zero. This includes the convection terms on the left hand side of equation 2.3 and the derivatives of macroscale flow stresses, which are the first 3 terms on the right hand side of equation 2.3. Once the flow is Reynolds averaged, these terms are invariant from REV to a neighboring REV. While the elimination of these terms from equation 2.3 is strictly valid for periodic porous medium geometries (example: heat exchangers), they may also be eliminated for other homogeneous porous media (example: canopy flows) if the size of the REV is large enough to minimize the influence of microscopic variations in the solid obstacle geometry. However, if the flow field is inhomogeneous, such as in the case of inhomogeneous geometry (by inhomogeneous, we mean that the REV geometry is not copied across the entire porous medium), macroscale flow gradients, or flow symmetry-breaking, the macroscale convection and stress gradients will also have to be modeled.

For the model problem (figure 1), the governing equation for macroscale momentum transport is,

$$\frac{\mu}{\Delta V} \int_{A_{interface}} n_j \, \partial_j \overline{u}_i \, dS - \frac{1}{\Delta V} \int_{A_{interface}} n_i \overline{p} \, dS + \rho \varphi g_i = 0 \qquad (2.5)$$

Note that the Reynolds stress terms will be zero due to the periodicity of the REV and the zero value of Reynolds stress at solid walls. The applied pressure gradient ($\rho g$) is either user-specified for periodic flows or derived from the macroscale pressure gradient in inhomogeneous flows. Therefore, the CNN model must predict the viscous and pressure drag forces in the first 2 terms that arise from the microscale flow distribution for the closure of equation 2.5. The sum of the pressure and viscous drag forces is often predicted using the Darcy-Dupuit-Forchheimer model for laminar flows as discussed in section 1. Since the solid obstacle surface is immersed in the REV, the CNN model used in the present work cannot accurately predict the stress distribution on the solid obstacle surface. This is because the no-slip boundary conditions at the solid obstacle surface are not enforced by the CNN model. Instead, the CNN model will predict the distribution of flow variables such as the microscale velocity or pressure in the entire volume of the REV. The divergence theorem is used to evaluate the drag force components from the volumetric distribution of the flow variables using equation 2.6.

$$\frac{\mu}{\Delta V}\int_{\Delta V}\frac{\partial}{\partial x_j}\left(\frac{\partial u_i}{\partial x_j}\right)dV - \frac{1}{\Delta V}\int_{\Delta V}\frac{\partial \bar{p}}{\partial x_i}dV + \rho\varphi g_i = 0 \quad (2.6)$$

In section 3, we evaluate two approaches to predict the microscale drag force components using the CNN model: (1) by modeling the velocity and pressure distributions, (2) by modeling the densities of the momentum sources of pressure and viscous drag. The densities of the momentum sources of viscous and pressure drag forces are the integrands under the volume integrals in the first two terms of equation 2.6, respectively.

## 3. Results and Discussion

### 3.1 Simulation of the turbulent flow dataset

Following the procedure described in section 2.1, we simulated turbulent flow in porous media for numerous values of porosity and compiled a dataset of the microscale distributions of the velocity and pressure. We interpolate the unstructured turbulent flow data to a structured grid by using bilinear interpolation in section 3.2 and a novel conservative interpolation method in section 3.3. This dataset represents two different flow regimes with respect to the porosity: low and intermediate porosity flow regimes. The low porosity flow regime ($\varphi < 0.8$) is characterized by recirculating vortices sandwiched between two strong shear layers around the solid obstacle surface (figure 3(a)). The intermediate porosity flow regime is characterized by vortex shedding and the formation of the von Karman instability (figure 3(c)). However, the pore size at intermediate porosity is small enough to cause a strong interaction between the flow around one solid obstacle and the neighboring solid obstacle. The high porosity flow regime ($\varphi > 0.95$) is not encountered in the present work. At high porosity, the solid obstacle size is small when compared to the pore size such that there is only a weak interaction between the flow around a solid obstacle and the neighboring solid obstacle. As a result, the drag force does not change significantly in this flow regime with respect to the porosity. It is evident in figure 3(b) that the viscous drag force per unit volume of the REV asymptotically approaches zero, especially when its magnitude is compared to the pressure drag. The primary source of viscous drag in porous media is the high shear caused by the constrained flow channels formed in between the solid obstacle surfaces. At high porosity, the solid obstacles do not channel the flow in between them, which causes the viscous drag magnitude to diminish.

In the present discussion, we have chosen to present three cases ($\varphi = 0.4$, 0.75, and 0.85 or cases 4, 27, and 47) as representative cases to demonstrate the sources of error and facilitate the discussion of the model results at the microscale level. The microscale flow distributions of velocity and pressure for all of the cases are shown in the online supplementary material (figures S1-S2). At $\varphi = 0.4$, the low porosity flow regime is encountered where both the pressure and viscous drag force components are significant. At $\varphi = 0.75$, the flow behavior is near the boundary between the low and intermediate porosity flow regimes. Even though recirculating vortices are encountered in this case with high stagnation pressure at the vertices of the square geometry, the pressure drag is 3 times the viscous drag because of diminishing shear stress in between the solid obstacle surfaces. At $\varphi = 0.85$, the intermediate porosity regime is encountered, which is characterized by shedding vortices and strong interaction of the vortex wake behind one solid obstacle and its downstream neighbor. This causes an increase in the pressure drag such that its magnitude is higher at $\varphi = 0.85$ when compared to $\varphi = 0.75$.

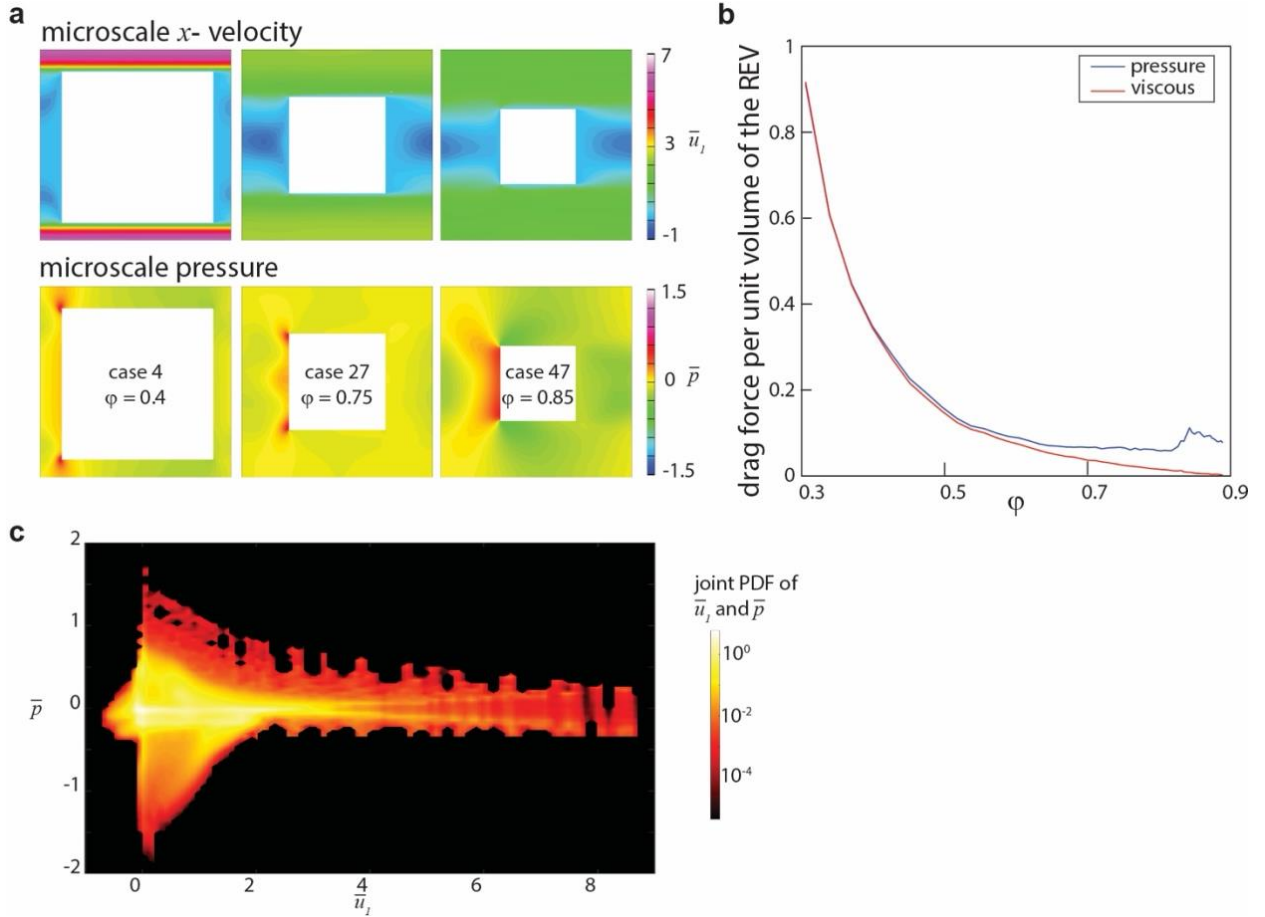

Figure 3: (a) the microscale *x*- velocity and pressure distributions for cases 4, 27, and 47 representing the low and intermediate porosity flow regimes that the CNN model must learn to predict. (b) the variation of the drag force components with respect to porosity. (c) the ranges of *x*- velocity and pressure distributions at the microscale level shown using a joint probability density function.

*3.2 Selection of the loss function for multivariate prediction*

Considering the range of flow behaviors observed in the turbulent flow dataset in our model problem, the first step in the development of the CNN model described in section 2.2 is to develop the capability to predict multiple flow variables as the output of the CNN model. Based on equation 2.6, we select the *x*- velocity and the pressure as the outputs of the CNN model so that we can compute their derivatives and estimate the drag force components. This is not trivial since the microscale *x*- velocity and pressure values in our dataset are not uniformly distributed. The joint probability density function of the microscale *x*- velocity and pressure (figure 3(c)) reveals that the range of the *x*- velocity is one order of magnitude greater than the range of the pressure. Additionally, there exists an inverse relation between the pressure and velocity in several flow scenarios. For example, flow stagnation points on the solid obstacle surface are characterized by high pressure and zero *x*- velocity. Whereas the flow in the channel regions in between the solid obstacles is characterized by low pressure and high *x*- velocity.

When the Mean Squared Error (MSE) loss function is used to train the CNN model, the non-uniform distribution of data in our samples introduces significant error for multivariate prediction of both *x*- velocity and pressure. The resulting model predicts the *x*- velocity, a variable with larger range, more accurately than the pressure, a variable with smaller range (figure 4). This is because the error in the prediction of the pressure is outweighed by the error in the prediction of *x*- velocity. Data normalization can be used to scale the dataset to have a zero mean and unit variance, but this introduces the need for additional model constants to recover the velocity and pressure from the scaled CNN model predictions. Data normalization is unsuitable for the present regression problem where the outputs are fluid dynamic variables that quantitatively estimate the drag force.

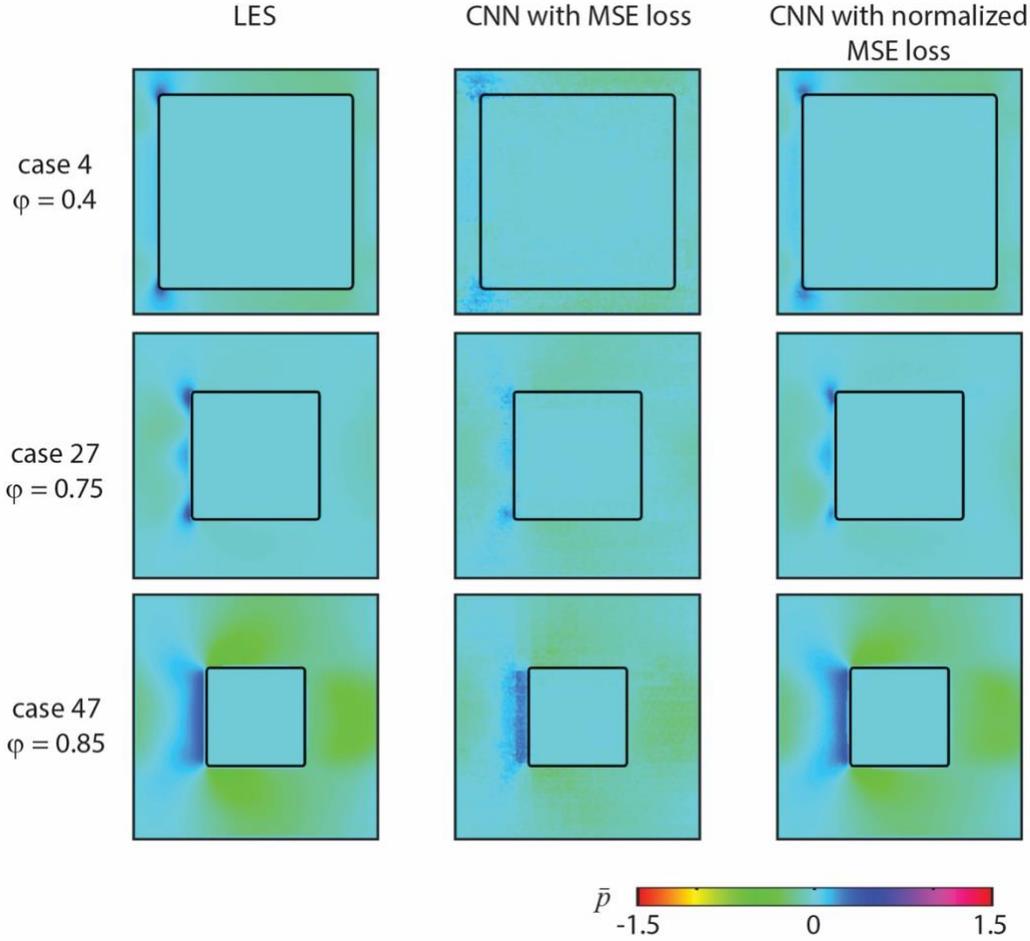

Figure 4: The normalized MSE loss function used in the present work (equation 3.1) provides better CNN model prediction accuracy when compared to the MSE loss function. The colored contours show the microscale pressure distribution for cases 4, 27, and 47.

To solve this problem, we are proposing to modify the loss function used to train the CNN model to penalize the optimization algorithm based on the normalized error calculated for each output variable. Therefore, the normalization procedure is performed during the model optimization rather than for the dataset, which remains in the original non-dimensional format that was simulated with LES. We have trained our CNN model to minimize the custom loss function – weighted normalized MSE (equation 3.1).

$$Loss = \frac{1}{n_{output}} \sum_{i=1}^{n_{output}} \frac{\sum_{j=1}^{n_{elements}}(\phi_{i,j,true}-\phi_{i,j,predicted})^2}{\sum_{j=1}^{n_{elements}}(\phi_{i,j,true})^2} \qquad (3.1)$$

where $n_{elements}$ is the number of elements in the output tensor for each variable, true values are simulated by LES, and predicted values are produced by the CNN model. In this loss function, the mean squared absolute error for each variable is normalized with respect to its mean value and then averaged with equal weighting. Equal weighting ensures that each variable contributes equally towards the calculation of the loss function. Since some fluid dynamic variables ($y$- velocity) may produce a zero mean over the REV, a small bias ($1 \times 10^{-20}$) is added to ensure the normalization procedure does not result in division by zero. When the CNN model is optimized by using the normalized MSE loss function (equation 3.1), the normalized MSE in the prediction of the microscale pressure decreases by 4 times when compared to the CNN model optimized by using the MSE loss function. The improvement in the CNN model prediction is evident in figure 4 where the model trained using normalized MSE loss function eliminates noise in the output variables, which is present for the model trained using MSE. Considering all of the validation cases in the dataset, the CNN model trained using the normalized MSE loss function has 4.9% normalized MSE and 2.5% macroscale error when compared to LES.

*3.3 Minimization of interpolation error for accurate prediction of the drag force components*

The goal of the CNN model framework is to estimate the drag force components that are acting on the solid obstacle surface in the porous medium. This requires the evaluation of volume integrals where the integrand has the first derivative of pressure and the second derivative of velocity (equation 2.6) for calculating the pressure and viscous drag components, respectively. In section 3.2, we have developed a CNN model that can predict the pressure and $x$- velocity distributions with less than 5% MSE at the microscale level. However, the estimation of the drag force components by using the CNN model output and equation 2.6 resulted in inaccurate predictions when compared to the drag force components simulated by LES (figure 5). The error in the prediction of the pressure drag in figure 5(b) is caused by the bilinear interpolation method used to interpolate the $x$- velocity and pressure distributions from the unstructured grid used for LES to the structured grid used for the CNN model. It is worth noting that the CNN model accurately predicts the interpolated LES result that was used to train and validate the model such that the two drag curves are virtually coincident in figure 5(b).

This presents a serious challenge in the application of CNN models to predict turbulent flow variables since unstructured grids with varying grid sizes are commonly used to simulate the flow. Typically, CNN models are designed to process inputs and outputs of prescribed dimensions. Interpolation of data is crucial to train models with a wide range of available data that may be derived from different CFD techniques and experiments. A good CNN model framework should be compatible with different types of data (structured, unstructured, and scattered) so that the model can be robustly trained on an expansive dataset derived from multiple sources. We investigated two possible solutions to increase the accuracy of the CNN model in predicting the drag force by modifying the interpolation method.

In the first solution, we continued to use the bilinear interpolation method and interpolated the LES flow data on to a structured grid with high grid resolution. The high grid resolution will minimize the error in the interpolation method, especially in the regions close to the solid obstacle surface where the pressure and velocity distributions vary significantly. In this approach, the unstructured LES data is interpolated to a 1280x1280 structured grid. It is not practical to train a CNN model with an output grid size of 1280x1280 since it substantially increases the number of trainable parameters in the model. Increase in the number of trainable parameters increases the computational requirements to train and evaluate the CNN model. The memory and processing requirements for high resolution data is especially high when larger and three-dimensional flow datasets are considered. Therefore, we upscaled the structured data from a 1280x1280 grid to a 128x128 grid. The upscaling procedure consists of the following steps: (1) the pressure gradient stored in the fine grid is integrated over the volume of the coarse cell, (2) the integral value of the pressure gradient is then divided by the volume of the coarse grid cell, and (3) the upscaled pressure gradient is stored on the coarse grid level. The upscaling procedure is similar to the LES spatial filtering operation. When the upscaled pressure gradient distribution is predicted as the output of the CNN model, the error in the pressure drag calculation decreases such that the average error across all 61 cases compared to LES is 10.4% (figure 5(c)). This is a significant improvement compared to the case of coarse grid bilinear interpolation, which had an average error of 101% in the prediction of pressure drag and resulted in the unphysical prediction of negative drag force.

However, fine grid interpolation does not eliminate the fundamental issue with the interpolation procedure – inaccurate approximation of conserved quantities. One drawback of the upscaling approach is the high magnitude of interpolation error that occurs at low values of porosity where the source of pressure drag is concentrated at the vertices of the square geometry of the solid obstacle. Resolving this issue would require more refined grids. Another drawback with the use of upscaling is that it does not accurately predict the viscous drag force. This is because the viscous drag force is calculated from the second derivative of velocity, which could not be approximated accurately with bilinear interpolation even with fine grids.

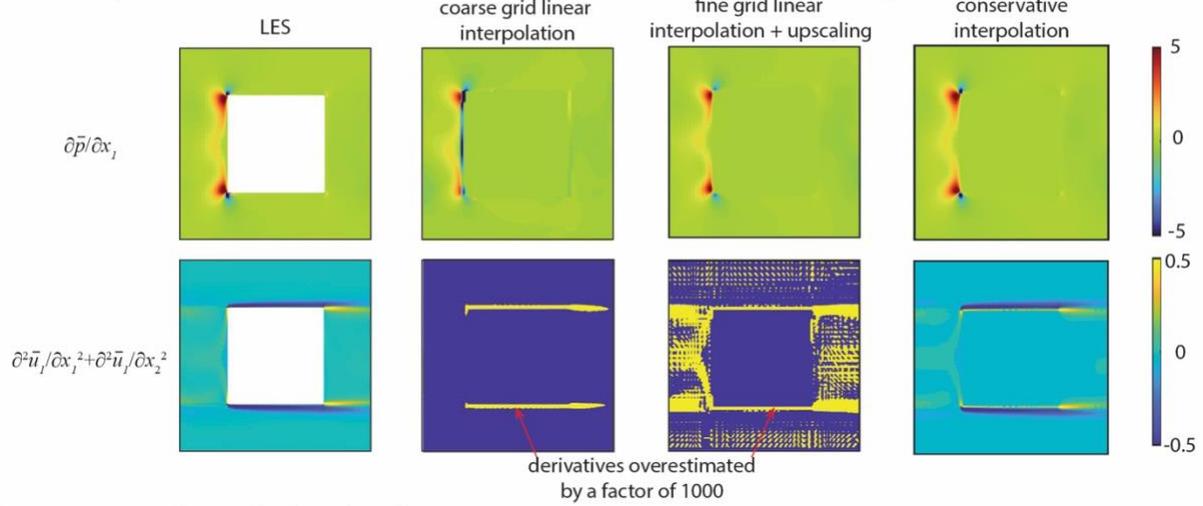

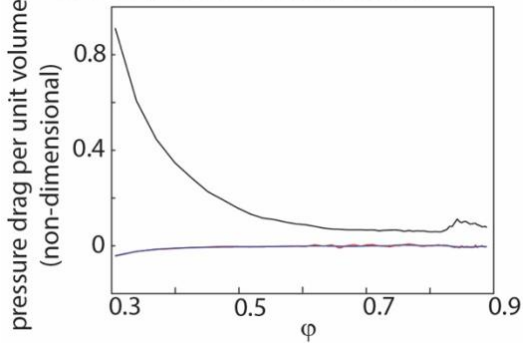
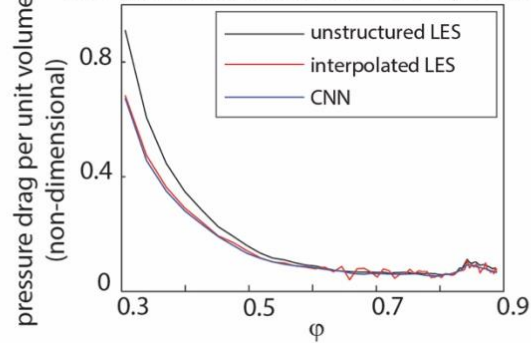

Figure 5: (a) A comparison of the calculation of the first derivative of pressure and the second derivative of velocity for different interpolation methods shows that linear interpolation is unsuitable for turbulent flow data stores on stretched unstructured grids. (b) As a result, the pressure drag values have high magnitude of error, which can be overcome by using a fine grid for linear interpolation followed by upscaling (c).

To overcome these drawbacks, we propose the second solution, which eliminates the interpolation error by ensuring that the governing equation (equation 2.6) is satisfied for the interpolated data used to train the CNN model. In equation 2.6, the pressure and viscous drag forces are calculated by integrating the gradients of pressure and shear stress components over the entire volume of the REV, respectively. Previously, the velocity and pressure distributions were predicted as the output of the CNN model. The drag force was calculated from the CNN model output in two steps – (1) the calculation of the gradient of the Cauchy stress tensor along the $x$- axis (flow direction) and (2) the volume integration of the gradient of the Cauchy stress tensor over the entire REV. Since there is no additional benefit in predicting the $x$- velocity and pressure distributions using the CNN model, it would be more beneficial to directly predict the distributions of the integrands in equation 2.6. This means that step (1) is performed before training the CNN model and the resulting stress gradient distributions are modeled as the output of the CNN model. The integrands in equation 2.6 can be interpreted as the densities of the momentum sources of pressure and viscous drag, respectively. Linear interpolation of the integrands from the unstructured CFD grid to the structured CNN grid introduces substantial error due to the concentration of flow stresses near the solid wall.

A conservative interpolation scheme is devised to interpolate the pressure and shear stress gradients from the unstructured CFD grid to the structured CNN grid to ensure that equation 2.6 is satisfied by the interpolated data on the structured CNN grid (figure 6). The conservative interpolation of variable ψ for a single structured grid cell is performed as per the equation 3.2.

$$\psi_{interpolated,i} = \frac{1}{\Delta V_i} \int_{\Delta V_i} \psi_{unstructured} dV \qquad (3.2)$$

where $\psi_{interpolated,i}$ is the value of ψ in the structured grid cell $i$, $\Delta V_i$ is the volume of structured grid cell $i$, and $\psi_{unstructured}$ is the value of ψ in the unstructured CFD grid. This conservative interpolation procedure is similar to volume-weighted interpolation since the governing equation includes only volume integrals.

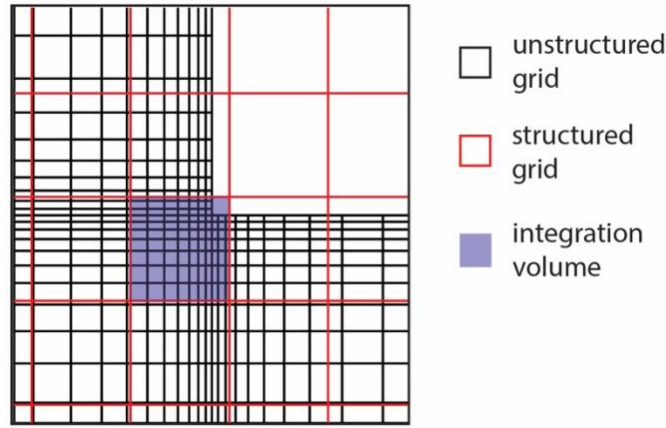

Figure 6: Schematic of the conservative interpolation from the unstructured CFD grid (black) to the structured CNN grid (red).

The implementation of the conservative interpolation scheme completely eliminated the interpolation error in the calculation of both the pressure and viscous drag forces using the LES data (figure 7). Subsequently, the CNN model is trained using the interpolated LES data to predict the densities of the momentum sources of pressure drag and viscous drag. Note that the CNN model is 'softly' constrained to satisfy the conservation of momentum (equation 2.6) by minimizing the normalized MSE loss function for the training dataset prepared using conservative interpolation. In this context, soft constraint means that the conservation law is satisfied by optimizing the model to fit the conserved quantities in the flow dataset, but it is not hardwired in the model. The resulting estimates of the pressure and viscous drag forces have 8% and 5.7% absolute error, respectively. The error in the prediction of the pressure drag force is higher than that of the viscous drag because the sources of pressure drag are concentrated at the vertices of the square solid obstacle, whereas the sources of the viscous drag are distributed near the solid obstacle surface. The performance of the CNN model may be improved when the solid obstacle geometry inherently distributes stresses along the surface, such as in the case of circular solid obstacles.

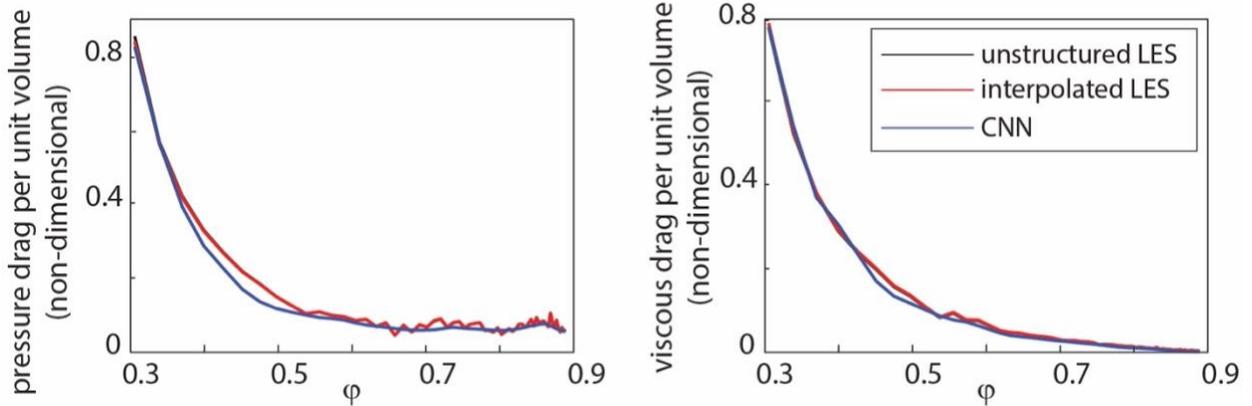

Figure 7: Pressure and viscous drag forces estimated by the CNN model after using the conservative interpolation scheme to prepare the training dataset.

## 4. Summary

In this paper, we have developed a convolutional neural network (CNN) model framework to predict the microscale turbulent flow distribution in periodic porous media. The microscale geometry of the solid obstacles is provided as the input to the CNN model using the signed distance function. The CNN model then learns to map the geometry of the solid obstacles to the output of the CNN model, which is the microscale flow distribution, from a Large Eddy Simulation (LES) dataset. We have then used the microscale flow distribution to estimate the pressure and viscous drag force components that act on the solid obstacle surface for macroscale model closure. This data-driven modeling procedure is divided into two steps: (1) the simulation of the turbulent flow dataset using LES, and (2) training and validation of the CNN model.

To develop the CNN model framework, we chose to simulate the dataset for a model problem of turbulent flow in periodic porous media composed of cylindrical solid obstacles with a square cross-section. We varied the porosity of the porous medium in the range 0.3 to 0.88 to generated 61 samples. The samples include a wide range of flow behaviors encompassing two unique flow regimes: the low and intermediate porosity flow regimes, which are characterized by recirculating and shedding vortex structures respectively. We note that training the CNN model for practical applications will require a more robust dataset that considers different solid obstacle geometries and Reynolds numbers. Future work must address the prohibitive computational requirements for generating a robust turbulent flow dataset for porous media flows. However, the validation of the CNN model trained with the present dataset revealed the following challenges hindering the development of macroscale closure models with this approach.

The first challenge is introduced by the need to predict multiple variables (pressure and velocity) in order to calculate the pressure and viscous drag force components. For turbulent flow in porous media, the values of microscale pressure and velocity inside the pores span different ranges, such that the range of $x$- velocity values is an order of magnitude larger than the range of the pressure values. Therefore, the CNN model trained using the popular Mean Square Error (MSE) loss function results in the model learning to accurately predict only the $x$- velocity distribution since the magnitude of $x$- velocity is typically larger than that of the pressure. To overcome this, we used normalized the loss function used to train the CNN model such that each component of the CNN model output is normalized with respect to its mean value. We demonstrated the use of the normalized MSE loss function in our CNN model decreased the model error in the simultaneous prediction of pressure and $x$- velocity by a factor of 4.

The second challenge is introduced by the need to interpolate the turbulence dataset from an unstructured grid used for LES to a structured grid of a specified size that is compatible with the CNN model. The interpolation error is compounded when the first and second derivatives of pressure and velocity are calculated to compute the drag force. Significant interpolation error is introduced by linear interpolation methods, especially in the near-wall regions where the LES grid is stretched to resolve the turbulent boundary layer. We evaluated two possible solutions to overcome this challenge. In the first solution, the error caused by the linear interpolation method is minimized by using a fine structured grid for interpolation and then upscaling the interpolated flow distribution to a coarse grid that is compatible with the CNN model. This approach decreased the error in the prediction of pressure drag by a factor of 10 (from 101% o 10.4%) when compared to the coarse grid linear interpolation. However, the estimation of the viscous drag continued to have substantial interpolation error. We addressed this in the second solution by using a conservative interpolation scheme that satisfies the macroscale governing equation of momentum on both the unstructured LES grid and the structured CNN grid. By using the conservative interpolation approach, we decreased the interpolation error in the CNN model to virtually zero for both the pressure and viscous drag calculations.

Therefore, we have developed a CNN model framework that processes turbulent flow data simulated by LES and then trained the model to predict the density of the momentum sources of the pressure and viscous drag forces. By using the proposed conservative interpolation method and the normalized MSE loss function, the resulting CNN model predicts the pressure and viscous drag forces with less than 10% mean absolute error over the range of porosities from 0.3 to 0.88. The CNN model offers a speedup in the calculation of the drag force components of $O(10^6)$ when compared to LES, where the CNN model takes less than 0.4 seconds on a desktop computer and the LES took 2 months of computation time on a linux cluster. Further development of the CNN model for turbulent flow in porous media with a more robust dataset holds tremendous potential for the design and modeling of optimized porous materials.

**Acknowledgements.** The authors acknowledge the computing resources provided by North Carolina State University High Performance Computing Services Core Facility (RRID:SCR_022168). AVK acknowledges the support of the Alexander von Humboldt Foundation through the Humboldt Research Award.

**Funding.** This research was funded by the National Science Foundation award CBET-2042834

**Declarations of interests.** The authors report no conflict of interest.

**Author ORCIDs.** V. Srikanth, https://orcid.org/0000-0002-2521-3323; A. V. Kuznetsov, https://orcid.org/0000-0002-2692-6907

**Appendix A: Details of the numerical model used to simulate the microscale turbulent flow dataset**

*A.1. Details of the Large Eddy Simulation model*

The governing equations for the LES are the filtered Navier-Stokes equations (A.1-A.2). The tilde ($\tilde{u}$) denotes the spatial filtering operator. Since periodic boundaries are used, the pressure variable $\tilde{p}$ is the filtered periodic pressure. To calculate the static pressure, we take the sum of the periodic pressure and the linearly varying pressure from the applied pressure gradient. The subgrid velocity scale is estimated using the Dynamic One-equation Turbulence Kinetic Energy (DOTKE) subgrid scale model (A.3-A.8). The subgrid scale filter length $\Delta$ is calculated as the cube root of the cell volume. Kim (2004) has demonstrated the model accuracy while using the cube root of the cell volume as the filter width for simulating the flow inside channels and around square cylinders with unstructured stretched grids. The characteristic subgrid length scale for the calculation of subgrid turbulent viscosity in equation A.4 is estimated as $\Delta$. A dynamic procedure is used to estimate the model constants $C_k$ and $C_\varepsilon$ (Kim and Menon 1997) where the grid scale velocity field is filtered to a second, test scale velocity field. The test filter length $\hat{\Delta}$ is set equal to $2\Delta$ (ANSYS Inc. 2016) and the model constants are determined by invoking similarity between the stresses at the two scales. In this procedure, the value of $C_k$ is limited by $-\mu/(k_{SGS}^{1/2}\Delta)$ to avoid a negative total viscosity. The Finite Volume Method (FVM) is used to solve the filtered governing equations and the DOTKE subgrid model. Note that a box LES subgrid scale filter is implicitly applied by the computational grid in the FVM.

$$\frac{\partial \widetilde{u_j}}{\partial x_j} = 0 \tag{A.1}$$

$$\frac{\partial \rho \widetilde{u_i}}{\partial t} + \frac{\partial \rho \widetilde{u_i}\widetilde{u_j}}{\partial x_j} = -\frac{\partial \widetilde{p}}{\partial x_i} + \frac{\partial}{\partial x_j}\left[(\mu + \mu_{T,SGS})\left(\frac{\partial \widetilde{u_i}}{\partial x_j} + \frac{\partial \widetilde{u_j}}{\partial x_i}\right)\right] + \rho g_i \tag{A.2}$$

$$\frac{\partial k_{SGS}}{\partial t} + \frac{\partial (\widetilde{u_j} k_{SGS})}{\partial x_j} = \left[C_k k_{SGS}^{\frac{1}{2}} \Delta \left(\frac{\partial \widetilde{u_i}}{\partial x_j} + \frac{\partial \widetilde{u_j}}{\partial x_i}\right)\right]\frac{\partial \widetilde{u_i}}{\partial x_j} - C_\varepsilon \frac{k_{SGS}^{\frac{3}{2}}}{\Delta} + \frac{\partial}{\partial x_j}\left(\mu_{T,SGS}\frac{\partial k_{SGS}}{\partial x_j}\right) \tag{A.3}$$

$$\mu_{T,SGS} = C_k k_{SGS}^{\frac{1}{2}} \Delta \tag{A.4}$$

$$\tau_{ij} = -2C_k k_{SGS}^{\frac{1}{2}} \Delta \widetilde{S_{ij}} + \frac{2}{3}\delta_{ij} k_{SGS}, L_{ij} = -2C_k k_{test}^{\frac{1}{2}} \hat{\Delta} \widehat{\widetilde{S_{ij}}} + \frac{1}{3}\delta_{ij} L_{kk} \tag{A.6}$$

$$C_k = \frac{1}{2}\frac{L_{ij}\sigma_{ij}}{\sigma_{ij}\sigma_{ij}}, \sigma_{ij} = -\hat{\Delta} k_{test}^{\frac{1}{2}} \widehat{\widetilde{S_{ij}}}, k_{test} = \frac{1}{2}\left(\widehat{\widetilde{u_k}\widetilde{u_k}} - \widehat{\widetilde{u_k}}\widehat{\widetilde{u_k}}\right) \tag{A.7}$$

$$C_\varepsilon = \frac{\widehat{(\partial \widetilde{u_i}/\partial x_j)(\partial \widetilde{u_i}/\partial x_j)} - (\partial \widehat{\widetilde{u_i}}/\partial x_j)(\partial \widehat{\widetilde{u_i}}/\partial x_j)}{\left((\mu + \mu_{T,SGS})\hat{\Delta}\right)^{-1} k_{test}^{\frac{3}{2}}} \tag{A.8}$$

The convective terms in equations A.2-A.3 are discretized using the bounded second-order central scheme (Leonard 1991) and the diffusive terms are discretized using the second-order central scheme. The pressure and velocity variables are stored in staggered locations. The pressure is stored at the centroid of the face of the cell, while the velocity is stored at the cell center. The PISO algorithm is used to solve the momentum and pressure Poisson equations in a segregated manner (Issa 1986). The simulation is advanced in time using a second-order implicit backward Euler method. The simulations are performed using the commercial code ANSYS Fluent 16.0. A detailed description of the numerical method is available in the ANSYS Fluent Theory Guide (ANSYS Inc. 2016).

*A.2 Validation of the numerical model*

The CNN model used in this work is trained with a numerically simulated dataset without including any experimental data samples. In this section, we demonstrate that the numerical model used to simulate the dataset can reproduce the pressure distribution on the solid obstacle surface measured by Aiba *et al.* (1982) for an in-line arrangement of circular tubes. The small size of the tube bank used in the experimental study of Aiba *et al.* (1982) makes it possible to simulate the experimental setup in the validation study. The limitations of this validation study are that (1) the Reynolds number of the flow used in the experimental work is an order of magnitude larger than that of the turbulent flow dataset, and (2) cylindrical solid obstacles with circular cross-section are used in the experiment instead of the square cross-section used for the turbulent flow dataset. However, the use of a large Reynolds number for validation is beneficial to determine the performance of the LES subgrid model under the condition that the subgrid scales contribute a significant portion of the turbulence kinetic energy. An excellent agreement between the simulation and experiment at the high Reynolds number implies an even better simulation accuracy at low Reynolds numbers. This is because the subgrid flow properties are close to the subgrid model assumptions at low Reynolds numbers. The use of cylindrical solid obstacles with circular cross-section instead of square cross-section does not affect the model validity since the underlying flow physics of the problem is similar.

The geometry used for the validation simulation is shown in figure 8. The validation simulation models the experimental setup with a dimensionless tube spacing of 1.6 and a Reynolds number of 41,000 (as per equation 2.2). Note that all the lengths are non-dimensionalized using the tube diameter. The following approximations are made while modeling the experimental setup. The flow at the center of the tube span is modeled by introducing a periodic boundary condition in the *z*- direction. The approximation follows from the nearly constant velocity distribution in the middle of the channel for turbulent flow. The span of the periodic domain in the *z*- direction is two times the pore size. The turbulence two-point correlation width is less than the span of the domain. Sufficiently long entrance and exit sections to the test section are introduced such that the flow becomes fully developed. The entrance and exit sections of the computational domain are 30 times the channel width. We note that these approximations coupled with the coarse grid resolution used for the high Reynolds number flow may lead to a quantitative mismatch between simulation and experiment.

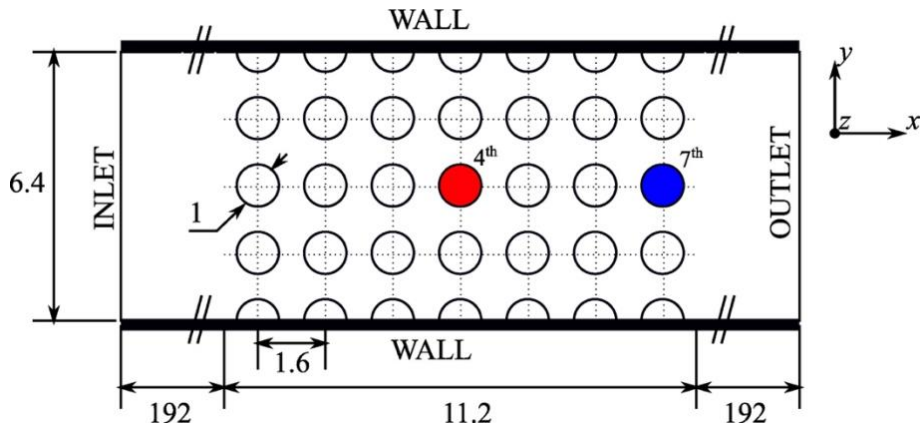

Figure 8: A sketch of the computational domain used to reproduce the experimental setup (Aiba *et al.* 1982) for validation of the numerical method. The simulations are performed to compare the coefficient of pressure on the colored tubes shown in the figure with that of the experiment. All the lengths are non-dimensionalized using the tube diameter.

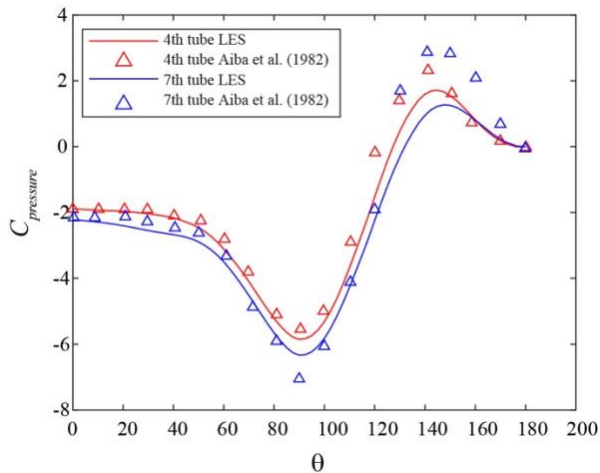

Figure 9: The distribution of the coefficient of pressure on the surfaces of tubes 4 (blue) and 7 (red) in the center row of the tube bank for the LES (solid line) and the experiment (triangle symbols).

The distributions of the coefficient of pressure ($C_{pressure}$) on the surface of the 4$^{th}$ and 7$^{th}$ tubes are used for comparison (figure 9). The $C_{pressure}$ is calculated as per the definition given by Aiba *et al.* (1982). The simulated $C_{pressure}$ distribution follows the same trend as that of the experiment. The simulated stagnation pressure is less than that of the experiment. In the low pressure regions, the quantitative agreement between the simulation and the experiment is excellent. The disparity between the simulation and the experiment is caused by the differences between the simulation and experimental setup, coarse grid resolution, and turbulence model limitations. However, these results suggest that the numerical method used in the present work is able to reproduce the flow behavior in porous media that is observed in experimental work. The numerical accuracy will improve further when compared to the validation case due to the high-resolution grids and the low Reynolds number used in the present work.

### *A.3 Grid Resolution study*

In the present work, we need to simulate a substantial number of cases using LES to train and evaluate the CNN model. Therefore, the grid resolution used in the study must find an acceptable compromise between resolving the relevant turbulent scales and predicting drag force accurately and the computational cost of simulating the dataset. We perform the grid resolution study for turbulent flow inside periodic porous media consisting of an in-line arrangement of circular tubes (similar to the validation study). We have set the Reynolds number of the flow equal to 1000, which is the same as the Reynolds number used for simulating the turbulent flow dataset. We have set the near-wall grid resolution equal to $0.001s$ (where $s$ is the pore size) so that the surface-averaged non-dimensional near-wall grid height ($\Delta y^+$) is consistently less than 1 for all the tested values of porosities ($\varphi$ = 0.5, 0.6, 0.72, and 0.8). In the center of the pore space, we have chosen three sizes for the grid cells ($\Delta x/s$ = 0.01, 0.02, and 0.03) to evaluate the adequacy of the grid resolution.

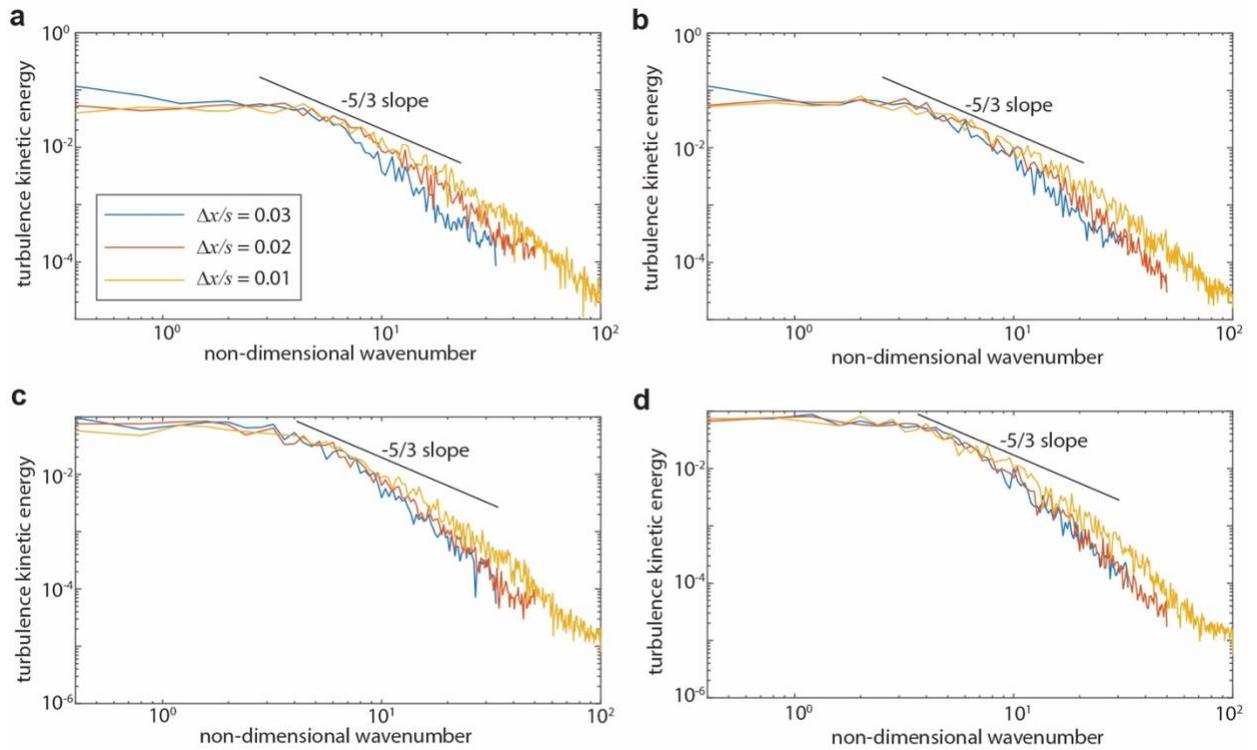

Figure 10: Wavenumber spectra of turbulence kinetic energy at different grid resolutions ($\Delta x/s$) for porosities (a) $\varphi$ = 0.5, (b) $\varphi$ = 0.6, (c) $\varphi$ = 0.72, (d) $\varphi$ = 0.8. The wavenumber is scaled with respect to $s$.

| Change in the grid resolution | φ = 0.5 | φ = 0.8 |
|---|---|---|
| Δx/s = 0.03 to 0.02 | 3.2% | 2.2% |
| Δx/s = 0.02 to 0.01 | 0.6% | 1.2% |

Table 1: Percentage change in the total drag force magnitude due to grid refinement

The resulting turbulence kinetic energy spectra (figure 10) reveal that all three grid resolutions resolve the energetic large scale turbulent eddies. Due to the confined space inside the pores, the large scale eddy subrange is followed immediately by the dissipative subrange. When we compare the energy spectra to the (-5/3) slope line, there is virtually no coincidence indicating the absence of the inertial subrange. This is also supported by the physiological explanation that the turbulent eddies inside the porous medium are always in proximity to production and dissipation. Significant deviations in the energy spectra for the different grid resolutions only emerge at the dissipative scales, which are characterized by low turbulence kinetic energy. While Direct Numerical Simulation of all the scales up to the Kolmogorov length scale would produce the most accurate results, we must strike a compromise between accuracy and computational cost in the present work. To determine the adequacy of the grid resolution in the present work, we compare the Reynolds averaged drag forces obtained from the simulations at different grid resolutions. The objective of the CNN model is to estimate the Reynolds averaged drag force. Since CNN model error is typically of the order of 10%, we assumed that the drag force magnitude has converged during grid refinement when the improvement in the drag force estimation of the order of 1% or less. By using this threshold value of numerical error, we demonstrate that a grid resolution of Δx/s = 0.02 is adequate to simulate the turbulent flow dataset (table 1).

### A.4 Details of the computational grid

The computational grid (figure 11) used to simulate the turbulent flow inside the REV is constructed using a block-structured grid topology to include high grid resolution cells near the surface of the solid obstacles and resolve the near-wall turbulent boundary layer. Grid cells are concentrated at the vertices of the square geometry, which are the locations of peak pressure, and near the solid obstacle surfaces, which are the locations of high shear stress. Adequately resolving the flow stresses near the solid obstacle surface is vital for the accurate estimation of the pressure and viscous drag forces.

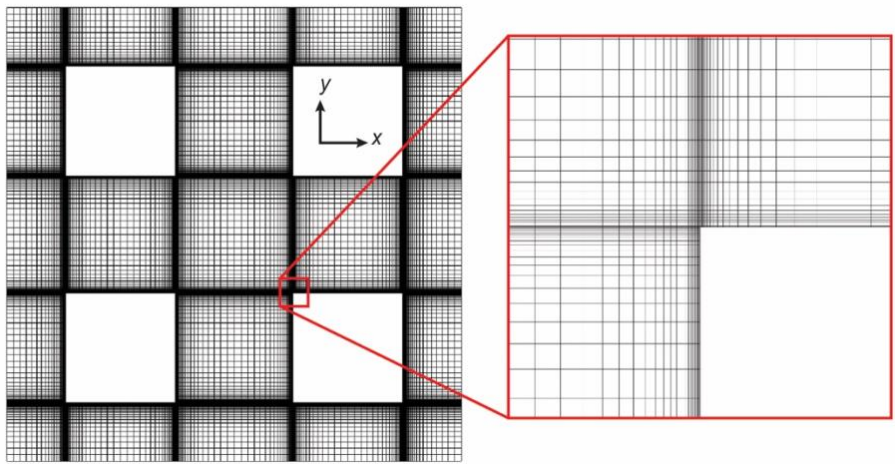

Figure 11: Computational grid used to simulate the turbulent flow inside the REV consisting of cylindrical solid obstacles of square cross-section.